\documentclass[aps,pre,onecolumn,superscriptaddress]{revtex4}
\usepackage{graphicx}
\usepackage{color}
\usepackage{amsmath}
\usepackage{amssymb}

\begin{document}

\title{Non-renewal resetting of scaled Brownian motion}

\author{Anna S. Bodrova}
\address{Humboldt University, Department of Physics, Newtonstrasse 15, 12489 Berlin, Germany}
%\address{Moscow Institute of Electronics and Mathematics, National Research University Higher School of Economics, 123458, Moscow, Russia}
%\address{Faculty of Physics, M.V.Lomonosov Moscow State University, Moscow, 119991, Russia}
\author{Aleksei V. Chechkin}
%\affiliation{Institute of Physics and Astronomy, University of Potsdam, 14476 Potsdam, Germany}
%\affiliation{Akhiezer Institute for Theoretical Physics, Kharkov Institute of Physics and Technology, Kharkov 61108, Ukraine}
\address{Institute of Physics and Astronomy, University of Potsdam, 14476 Potsdam, Germany}
\address{Akhiezer Institute for Theoretical Physics,\\ Kharkov Institute of Physics and Technology, Kharkov
61108, Ukraine}
\author{Igor M. Sokolov}
%\affiliation{Humboldt University, Department of Physics, Newtonstrasse 15, 12489 Berlin, Germany}
\address{Humboldt University, Department of Physics, Newtonstrasse 15, 12489 Berlin, Germany}

\begin{abstract}
We investigate an intermittent stochastic process, in which the diffusive motion with time-dependent diffusion
coefficient $D(t)\sim t^{\alpha-1}$,
$\alpha>0$ (scaled Brownian motion), is stochastically reset to its initial position and starts anew.
The resetting follows a renewal process with either exponential or power-law distribution of the waiting times
between successive renewals. 
The resetting events, however, do not affect the time dependence of the diffusion coefficient, so that the whole process appears to be a non-renewal one.  
We discuss the mean squared displacement of a particle and probability density function of its positions
in such a process.
We show that scaled Brownian motion with resetting demonstrates a rich behavior whose properties essentially depend 
on the interplay of the parameters of the resetting process and the particle's displacement in a free motion.
The motion of particles can
remain either almost unaffected by resetting, but can also get slowed down or even be completely suppressed.
Especially interesting are the nonstationary situations in which the mean squared displacement stagnates but the distribution of
positions does not tend to any
steady state. \color{black} This behavior is compared to the situation (discussed in the other paper of this series) in which the memory on the value of the diffusion coefficient at a resetting time 
is erased, so that the whole process is a fully renewal one. We show that the properties of the probability densities in such processes
(erazing or retaining the memory on the diffusion coefficient) are vastly different. \color{black}
\end{abstract}

\maketitle

\section{Introduction}

Resetting represents a class of stochastic processes, when a random motion is from time to time terminated and
restarted from given
initial conditions. The instant of restarting can depend on the state of the process (e.g. it may be restarted
under level crossing, like in many
neuronal models \cite{Gerstner}) or may be independent of that. This last class of the processes (motion under
stochastic resetting) is what we consider here.
One of the first studies of the processes with resetting was devoted to a discrete-time stochastic
multiplicative model
\cite{Manrubia99}. 

The random motion under stochastic resetting arises as the interplay of two distinct random processes: the
resetting process, a point process on the
real line representing the time axis, and particle's motion between the resetting events, which we will call
the
displacement process. The first work in the direction we follow in the present paper concentrated on the case
when the displacement process is an ordinary Brownian motion \cite{EvansMajumdar},
i.e. a Markovian process with stationary increments. The same is true for L\'evy flights considered in
\cite{levy1, levy2}, where, at difference to
Brownian motion, the trajectories of the free displacement process are discontinuous. 
Starting from the one-dimensional Brownian motion of single particles with resetting to the initial position
\cite{EvansMajumdar, MSS2015}, the
process was further generalized to two and higher dimensions \cite{high12}, \textcolor{black}{to motion
in a bounded domain with reflecting \cite{CS2015}
and adsorbing \cite{ArnabPal} boundaries and in an external potential \cite{P2015,expot}. Also the cases with
several choices of resetting position
\cite{EM2011, BS2014, MSS2015a}, with non-static restart points \cite{nonstatic} and for several interacting
Brownian particles with resetting \cite{brm2}
were discussed. Resetting has been investigated in the context of reaction-diffusion with stochastic decay
rates \cite{reacdiff} and the branching processes \cite{branch1, branch2}. Large deviations and phase transitions 
for Markov processes under resetting were considered in Ref. \cite{Rosemary}.}

Stochastic resetting of a diffusion process fundamentally changes its properties due to a competition between
the tendency of
diffusive spreading and repeated returns to the initial state. The ordinary normal diffusion process
interrupted at a constant rate by resetting to the
initial position \cite{EvansMajumdar} generates a non-equilibrium stationary state (NESS). However, the
limitation to constant resetting rate
severely restricts the applicability to memoryless resetting processes. The more general case of Gamma and
Weibull distributions of waiting times between the resetting events
was discussed in \cite{res2016}. Resetting with position-dependent resetting rate \cite{EM2011} and with 
time-dependent resetting rate \cite{palrt}, and resetting with power-law distribution of waiting times between
resetting events
\cite{NagarGupta} have been also considered. Resetting-induced NESS has also been studied in many-body systems
such as coagulation-diffusion processes \cite{coagdiff}.

Another important direction of work is connected with investigations on non-Markovian processes with
resetting. Thus, Ref. \cite{boyer2017} discusses resetting of a
particle to a position chosen from its trajectory in the past according to some memory kernel.
Another displacement process with memory considered corresponds to a continuous-time random walk with or
without drift \cite{MV2013, MC2016,
Sh2017,ctrw}.
 
In present paper we consider scaled Brownian motion (SBM) with stochastic resetting. SBM is a paradigmatic
Gaussian process governed by the overdamped Langevin equation with the diffusion coefficient which scales as a power law in time,
\begin{equation}
\frac{dx(t)}{dt}=\sqrt{2D(t)}\eta(t) ,
\end{equation}
where $D(t) \simeq t^{\alpha-1}$ with $\alpha > 0$. Here $\eta(t)$ represents white Gaussian noise with zero
mean $\left\langle \eta(t) \right\rangle
= 0$ and covariance $\left\langle \eta(t_1)\eta(t_2) \right\rangle = \delta\left(t_1-t_2\right)$. Setting
\begin{equation}
D(t)= \alpha K_{\alpha}t^{\alpha-1}, 
\label{Dt}
\end{equation}
one gets the mean squared displacement (MSD)
\begin{equation}\left< x^2(t) \right> = 2K_{\alpha}t^{\alpha} .
\label{x2-AD}\end{equation}
For $0<\alpha<1$ the motion is subdiffusive, and for $\alpha>1$ one speaks about superdiffusion
\cite{sokolovsub,sokbook,sokolovSM,MetzlerREVIEW,MetzlerPT,Bouchaudrev,zaburdaev,hoefling}. The case $\alpha =
2$ corresponds to ballistic spread, and
cases with $\alpha >2$ are termed superballistic or hyperdiffusive. In the limiting case $\alpha=0$ the
diffusion process is ultraslow with
logarithmic time dependence of the MSD \cite{ultraslow}. We note that the underdamped Langevin equation with time-dependent diffusion coefficient has been studied in
\cite{underdamped1,underdamped2}.

The SBM as a model for anomalous diffusion was first introduced by Batchelor to model turbulent dispersion
\cite{batchelor},
where the particles' spread is described by the Richardson's law \cite{Richardson} with the 
exponent $\alpha = 3$. Interestingly enough, the alternative models were the L\'evy flights (introduced long
before the name was coined), see Sec.
24.4 of Ref. \cite{Monin}, and L\'evy walks \cite{Shlesinger}.

SBM was used to describe fluorescence recovery after photo-bleaching in various settings \cite{saxton}, as
well as anomalous diffusion in various
biophysical contexts including brain matter \cite{novikov1, novikov2}. Time-dependent diffusion coefficient
may be observed in systems with
time-dependent temperature, such as melting snow \cite{molini,snow} or free cooling granular gases
\cite{brilbook,ggg1,ggg2}. Granular gas of
viscoelastic particles represents an illuminating example of a many-particle system where the self-diffusion
follows subdiffusive SBM with
$\alpha=1/6$; for a granular gas of particles colliding with constant restitution coefficient SBM with
$\alpha=0$ has been observed \cite{megg}.

The very term SBM was introduced in Ref.\cite{LimSBM}, where the authors compare the properties of SBM and
fractional Brownian motion (FBM). Both
processes are Gaussian random processes with the same single-time probability density functions, but are
intrinsically different in many other respects.
Thus, the SBM is a Markovian process with non-stationary increments, whereas the FBM is non-Markovian, but
possesses stationary increments.
%The SBM is a Gaussian process with non-stationary increments, in which respect it differs from L\'evy flights
(which always show
%superdiffusive behavior) and from fractional Brownian motion (FBM), which has both sub- and superdiffusive
regimes, see \cite{LimSBM}. Both these
%processes have stationary increments.
In contrast to FBM, the SBM exhibits discordance between its ensemble and time averaged MSDs, which is the
sign
of ergodicity breaking \cite{sokolovsub}. The non-ergodicity of SBM does not however go hand in hand with
strong difference between its different
realizations: its heterogeneity (ergodicity breaking) parameter tends to zero for long trajectories
\cite{Hadise}.

In the subdiffusive case the SBM can be considered as a mean-field approximation for the CTRW model
\cite{ThielSok} with a power-law waiting time
probability density function, which also has non-stationary increments. However, in SBM this non-stationarity
is modeled via the explicit time
dependence of the diffusion coefficient, while the CTRW, being of the renewal class, lacks explicit time
dependences of its parameters. On the other
hand, SBM is a Markovian process, while CTRW is a non-Markovian (semi-Markovian) one. Nevertheless, aging
properties of both processes are very
similar.
 
\color{black}
Therefore, just like in CTRW, two situations can be discussed: the dynamics of the underlying process 
is rejuvenated after resetting, or is not influenced by the resetting of the coordinate. In CTRW the first 
assumption would mean that a new waiting period starts immediately after the resetting event, see \cite{Sh2017} for the discussion
of the corresponding physical assumptions. In the second situation the waiting period started before the resetting event is not interrupted by the resetting.
Ref. \cite{Sh2017} concentrated on the first situation, corresponding to the renewal property of the whole process.

In the SBM the first assumption corresponds to the situation when the diffusion
coefficient also resets to its initial value, while the second situation corresponds to the case when only the position of the particle is
altered by the resetting events and the diffusion coefficient remains unaffected.
The two situations are quite different in their behavior. In the present work we concentrate on the second,
non-renewal, situation, while the first, fully renewal one, is considered in the other work of this series, Ref. \cite{Anna0}.
We analytically derive MSD and PDF for the cases of exponential and power-law resetting, and compare
our predictions with the results of numerical simulations. 
\color{black}

We proceed as follows. In the next Section II we define the main quantities describing the behavior of the
system with resetting and describe the details of numerical simulation.
In Sections III and IV we give the analytic results for SBM with exponential and power-law resetting,
correspondingly, and compare them with the numerical simulations.
Finally, we give our conclusions in Section V.

\section{Stochastic resetting}

Let us consider the particle returning to the initial position $x=0$ at random times. We denote by $\psi(t)$
the
probability density function (PDF) of waiting times between two consecutive resetting events. In the present
work we
concentrate on the two cases: the first one is when this PDF is exponential (which corresponds to a Poissonian
resetting process)
$\psi (t) \sim {e^{ - rt}}$, in the second one it follows a power law $\psi (t) \sim t^{-1-\beta}$.
The survival probability $\Psi(t)$ gives the probability that no resetting event occurs between 0 and $t$,
\begin{equation}
\Psi (t) = 1 - \int\limits_0^t {\psi (t')dt'}  = \int\limits_t^\infty  {\psi (t')dt'}\,.
\label{surv}
\end{equation}
Sometimes, especially for the case of the power-law PDF, it is convenient to switch between the time and the
Laplace domains.
The Laplace transform of the resetting PDF is
\begin{equation}
\tilde{\psi}(s)=\int_0^{\infty}\psi(t)\exp(-ts)dt\,.
\end{equation}
The Laplace transform of the survival probability can be expressed via $\tilde{\psi} (s)$ as
\begin{equation}
\tilde{\Psi} (s) = \frac{1-\tilde{\psi}(s)}{s}\,.
\label{survivalLaplace}
\end{equation}
The probability density $\psi_n(t)$ that the $n$-th resetting event happens at time $t$ satisfies the renewal
equation \cite{sokbook}
\begin{equation}
\psi_n(t) = \int_0^{t}\psi_{n-1}(t^{\prime})\psi(t-t^{\prime})dt^{\prime}\,,
\end{equation}
and the sum of all $\psi_n(t)$ gives the rate of resetting events at time $t$: 
\begin{equation}
\kappa(t)=\sum_{n=1}^{\infty}\psi_n(t)\,.
\end{equation}
Its Laplace transform yields
\begin{equation}
\tilde \kappa (s) = \sum\limits_{n = 1}^\infty {{{\tilde \psi }^n}(s)} = \frac{{\tilde \psi (s)}}{{1 - \tilde
\psi (s)}}\,.
\label{phis}
\end{equation}
The probability to find the particle at location $x$ at time $t$ (PDF) is 
\begin{equation}
p(x,t) = \Psi (t){p_0}(x,t,0) + \int\limits_0^t dt^{\prime} \kappa (t^{\prime})\Psi (t -
t^{\prime})p_0(x,t,t^{\prime})\,.
\label{eqprob1}
\end{equation}
Here the first term accounts for the realizations where no resetting took place up to the observation time
$t$. The weight of such realizations in the
ensemble of all realizations is given by $\Psi (t)$. The second term accounts for the case, when the last
resetting event before the observation
occurs at the time $t^{\prime}$ (the probability of which is $\kappa (t^{\prime})dt'$), no resetting occurs
between $t^{\prime}$ and $t$, and the particle
moves freely between these two instants of time. The first term may be safely neglected at long times
$t\to\infty$, and the PDF
of the particle's positions at such long times is
\begin{equation}
p(x,t) \simeq \int\limits_0^t dt^{\prime} \kappa (t^{\prime})\Psi (t - t^{\prime})p_0(x,t,t^{\prime})\,.
\label{eqprob}
\end{equation}
Between $t^{\prime}$ and $t$ the particle performs free SBM with PDF given by
\begin{equation}
p_0(x,t,t^{\prime})=\sqrt{\frac{1}{4\pi
K_{\alpha}\left(t^{\alpha}-t^{\prime\alpha}\right)}}\exp\left(-\frac{x^2}{4K_{\alpha}\left(t^{\alpha}-t^{\prime\alpha}\right)}\right)\,.
\label{p0sbm}
\end{equation}
%\Eq.~(\ref{eqprob1}) differs from the analogous equation for the case of Brownian motion obtained in
%\cite{EvansMajumdar} only by the fact
%that Eq.~(\ref{p0sbm}) is used as a diffusion propagator.

Multiplying Eq. (\ref{eqprob1}) by $x^2$ and performing the integration over $x$, we get the equation for the
MSD of particles:
\begin{equation}
\left\langle x^2(t)\right\rangle = 2K_{\alpha} t^{\alpha} \Psi(t) + 2K_{\alpha}\int_0^t dt^{\prime}
\kappa(t^{\prime})\Psi(t-t^{\prime})\left(t^{\alpha}-t^{\prime \alpha}\right)\,.
\label{x2}
\end{equation} 
At long times $t\to\infty$ the first term may be neglected and we obtain for the MSD
\begin{equation}
\left\langle x^2(t)\right\rangle \simeq 2K_{\alpha}\int_0^t dt^{\prime}
\kappa(t^{\prime})\Psi(t-t^{\prime})\left(t^{\alpha}-t^{\prime \alpha}\right)\,.
\label{x21}
\end{equation} 
The MSD may or may not be determined by the form of the PDF in the bulk and has to be calculated separately: A
very peculiar situation corresponding
to such a case, when the MSD stagnates but the bulk of the distribution shrinks, appears for power law waiting
time distributions with $1 < \beta < 2$.

In what follows we obtain the PDF, Eq. (\ref{eqprob1}), and the MSD, Eq. (\ref{x2}), for exponential and
power-law resetting waiting time densities
for long times analytically, and compare them to the results of numerical simulations. 
%\section{Numerical simulations}

The event-driven simulations are performed as follows. For a given sequence of the output times $t$ we
simulate the sequence of resetting events, find the time of
the last resetting event $t' < t$ and set $x(t)=0$. Then the position of the particle at time $t$ is
distributed according to a Gaussian
with zero mean and variance $\langle x^2 (t) \rangle = 2 K_\alpha (t^\alpha - t'^\alpha)$. The corresponding
Gaussian can be obtained from
a standard normal distribution generated using the Box-Muller transform. The results are averaged over
$N=10^4$ to $10^6$ independent runs.
In all our simulations $K_\alpha$ is chosen in such a way that $\alpha K_\alpha = 1$.

\section{Scaled Brownian motion with exponential resetting}

The simplest and most studied case corresponds to exponentially distributed  waiting times between resets
\begin{equation}
\psi (t) = r{e^{ - rt}}\,.
\label{pdfexp}
\end{equation}
In this case the resets occur at constant rate $r$. The survival probability, according to Eq. (\ref{surv}),
follows
\begin{equation}
\Psi (t) = {e^{ - rt}}\,.
\label{Psit}
\end{equation}
The rate at which resetting events follow is constant,
\begin{equation}
\kappa (t) = r\,.
\label{phir}
\end{equation}
This means that the resetting events occur with the same probability at any given interval $dt$ of time.

\subsubsection{Mean-squared displacement.} MSD for SBM with exponential resetting can be obtained by inserting
Eqs. (\ref{Psit}) and (\ref{phir}) into Eq. (\ref{x2}):
\begin{equation}
\left\langle x^2(t)\right\rangle =
2K_{\alpha}t^{\alpha}-2K_{\alpha}rt^{1+\alpha}e^{-rt}\frac{M\left(\alpha+1,\alpha+2,rt\right)}{\alpha+1}\,,
\end{equation}
where $M(a,b,z)$ is the Kummer function defined as \cite{as}
\begin{equation}
M(a,b,z)=\frac{\Gamma(b)}{\Gamma(a)\Gamma(b-a)}\int_0^1dte^{zt}t^{a-1}\left(1-t\right)^{b-a-1}
\end{equation}
with $\Gamma(z)$ being the Gamma function. 
%\begin{equation}\Gamma\left(z\right)=\int_0^{\infty}dt e^{-t}t^{z-1}\,.\end{equation}
Expanding the Kummer function $M\left(\alpha+1,\alpha+2,rt\right)$ for $rt\gg 1$ \cite{as},
\begin{equation}
M(\alpha+1,\alpha+2,rt) = \left(\alpha+1\right)\frac{e^{rt}}{rt}\left[ 1 + O \left(\frac{1}{rt}\right) \right]
,
\end{equation} 
we get the power-law dependence for MSD at long times
\begin{equation}
\left\langle x^2(t)\right\rangle \simeq \frac{2\alpha K_{\alpha}}{r}t^{\alpha-1}\,.
\label{x2exp}
\end{equation}
The exponent in the time-dependence of MSD is always by one smaller than in the case of free diffusive motion
without resetting. In such a way,
resetting affects SBM in a similar way as putting the particle performing SBM into a confining harmonic
potential \cite{MetzlerSBM}, or in fractional Brownian motion
in a fully renewal case \cite{Oshanin}.
For $\alpha=1$, we reproduce the result for standard Brownian motion with exponential resetting, namely
\begin{equation}
\left\langle x^2(t)\right\rangle = \frac{2 K_{1}}{r}\,.
\label{x2expobm}
\end{equation}
For ballistic motion between resetting events, $\alpha=2$, the motion with resetting shows the MSD behavior
akin to normal diffusion. The
superdiffusive SBM with $1<\alpha<2$ becomes subdiffusive in a presence of resetting. The most interesting
case corresponds to the subdiffusive SBM,
$0<\alpha<1$, when the MSD decays to zero following a power law. It means that due to slowing down of the
motion in the course of time the particle is
unable to get far away from the initial point between the resetting events, and tends to remain in the
vicinity of the origin.

\begin{figure}[htbp]
  \centerline{
\includegraphics[width=0.7\columnwidth]{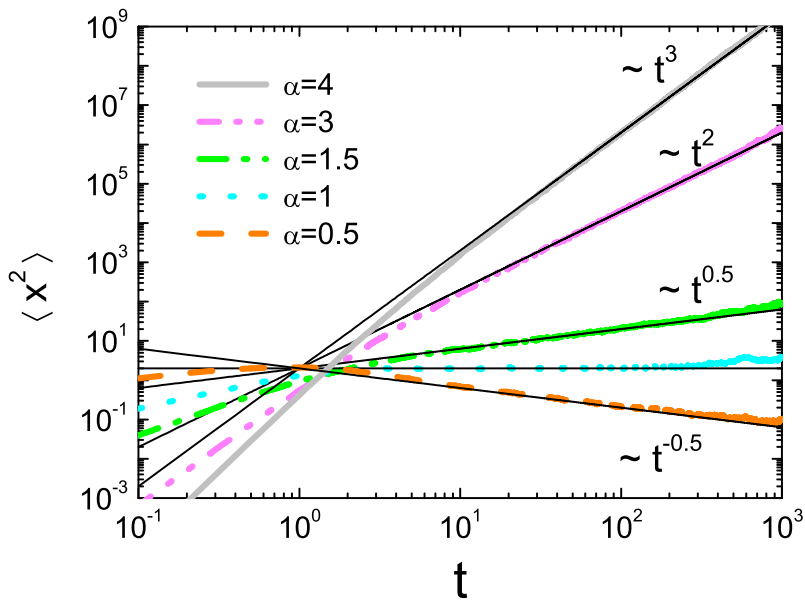}}
\caption{MSD for SBM with exponential resetting: theoretical results, Eq.~(\ref{x2exp}) (thin black solid
lines) and computer simulations
(thick colored lines) obtained for $\alpha = 0.5, 1, 1.5, 3, 4$.}
\label{GR2exp}
\end{figure} 

The analytical result for MSD in the case of exponential resetting, Eq.~(\ref{x2exp}), has been compared with
the numerical simulations showing the full agreement.
In Fig. \ref{GR2exp} we present the MSD for different values of $\alpha$. The thick colored lines correspond
to the numerical results, the thin solid lines corresponds to
asymptotics as given by Eq. (\ref{x2exp}). 
Light gray line corresponds to initially superdiffusive motion with exponent $\alpha=4$, which turns again to
superdiffusion, but with lower power exponent $\alpha-1=3$. For $\alpha=3$ initially
superdiffusive motion turns into ballistic motion with exponent $\alpha-1=2$ (magenta line in Fig.
\ref{GR2exp}). For $\alpha=1.5$ initially
superdiffusive motion turns into subdiffusion with $\alpha-1=0.5$ in the presence of resetting (green line).
In the case of the ordinary diffusion with $\alpha=1$
the MSD stagnates, as predicted in \cite{EvansMajumdar} (blue line). Initially subdiffusive motion with
$\alpha=0.5$ becomes trapped
in vicinity of the origin: the MSD tends to zero as $t^\gamma$ with $\gamma = \alpha-1=-0.5$ (orange line).

\subsubsection{Probability density function.} 
\begin{figure}[htbp]
  \centerline{
\includegraphics[width=0.7\columnwidth]{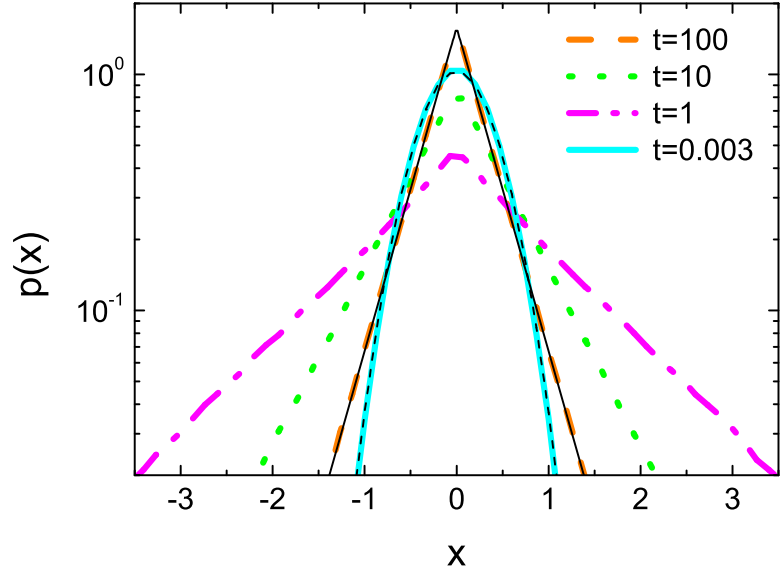}}
\caption{The PDF for subdiffusive SBM ($\alpha=0.5$) with Poissonian resetting for $t=0.3, 1, 10, 100$ (blue,
magenta, green and orange lines, correspondingly).
The width of PDF is first growing, and then starts to decrease. At very short time $t=0.003$ the PDF is
Gaussian (light blue line), reproducing the PDF of a free SBM (thin black dashed line).
At long times PDF is described by Eq. (\ref{pdfexp1}) (thin black solid line).
}
\label{GSBMexp}
\end{figure}

\begin{figure}[htbp]
  \centerline{
\includegraphics[width=0.7\columnwidth]{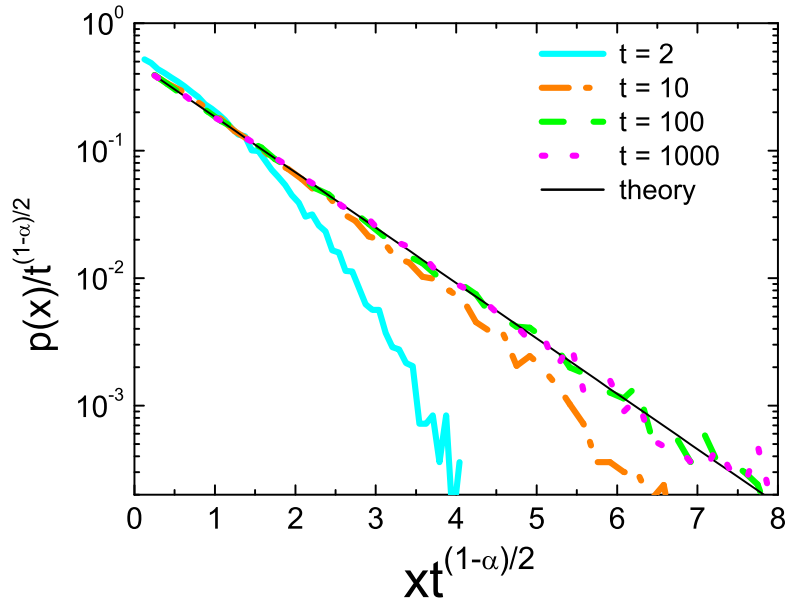}}
\caption{The PDF for superdiffusive SBM ($\alpha=3$) with Poissonian resetting as obtained in numerical
simulations for $t= 2, 10, 100$ and 1000 (thick colored lines),
and the prediction of the scaling form, Eq.(\ref{pdfexp1}). }. 
\label{GSBMexpsuper}
\end{figure}

Let us now obtain the asymptotic form of the PDF for SBM with
exponential resetting
valid in the long time limit for $t^{\alpha+1}\gg x^2/({K_{\alpha}r})$. 
Eq.~(\ref{eqprob}), together with Eqs. (\ref{p0sbm}), (\ref{Psit}) and (\ref{phir}), results in
%\begin{equation}p(x,t) = r\int\limits_0^t dt' e^{-r\left(t-t^{\prime}\right)}p_0(x,t,t^{\prime})\,.
%\label{eee}
%\end{equation} 
%Inserting the PDF $p\left(x,t,t^{\prime}\right)$ for free SBM, Eq. (\ref{p0sbm}), into Eq. (\ref{eee}), we
get
\begin{equation}\label{pxtexp}
p(x,t) \simeq
r\int_0^{t}dt'\exp\left(-\frac{x^2}{4K_{\alpha}\left(t^{\alpha}-t^{\prime\alpha}\right)}\right)\frac{\exp\left(-r\left(t-t^{\prime}\right)\right)}{\sqrt{4\pi
K_{\alpha}\left(t^{\alpha}-t^{\prime\alpha}\right)}}\,.
\end{equation}
Using new variable $\zeta=1-t^{\prime}/t$ we rewrite Eq.~(\ref{pxtexp}) as
\begin{equation}
p(x,t)\simeq rt\int_0^{1}d\zeta\frac{e^{\varphi(\zeta)}}{\sqrt{4\pi
K_{\alpha}t^{\alpha}\left(1-\left(1-\zeta\right)^{\alpha}\right)}}\,,
\label{bbb}
\end{equation}
where 
\begin{equation}
\varphi\left(\zeta\right)=-rt\zeta-\frac{x^2}{4K_{\alpha}t^{\alpha}\left(1-\left(1-\zeta\right)^{\alpha}\right)}\,.
\end{equation}
The major contribution to the integral Eq.(\ref{bbb}) comes from a small interval in the vicinity of
$\zeta_{\rm max}$, where $\varphi\left(\zeta\right)$
attains its maximum, which is given by the solution of $\varphi'(\zeta_{\rm max})=0$.
For $x$ fixed and $t$ large this maximum shifts closer and closer to zero, so that the approximation
\begin{equation}
1-\left(1-\zeta_{\rm max}\right)^{\alpha}\approx \alpha\zeta_{\rm max}
\end{equation}
holds, and $\zeta_{\rm max}$ can be estimated as
\begin{equation}
\zeta_{\rm max} \simeq \sqrt{\frac{x^2}{4\alpha K_{\alpha}r t^{\alpha+1}}}\,.
\label{zetamax}
\end{equation}
The integral, Eq.(\ref{bbb}), can then be evaluated using the standard Laplace method (i.e. expanding the
argument of the exponential up to the second order) thus
giving
\begin{equation}
 p(x,t) \simeq \frac{rt}{\sqrt{4\pi K_{\alpha}t^{\alpha}\alpha\zeta_{\rm max}}}e^{\varphi(\zeta_{\rm
max})}\int_{-\infty}^{\infty}d\zeta
e^{-\frac12\left(\zeta-\zeta_{\rm max}\right)^2\left|\varphi^{\prime\prime}\left(\zeta_{\rm
max}\right)\right|} =
\frac{rt e^{\varphi\left(\zeta_{\rm max}\right)}}{\sqrt{2K_{\alpha}t^{\alpha}\alpha\zeta_{\rm
max}\left|\varphi^{\prime\prime}\left(\zeta_{\rm
max}\right)\right|}}
\end{equation}
with $\varphi^{\prime\prime}\left(\zeta_{\rm max}\right)$ being the second derivative of $\varphi$ at its
maximum.
Performing calculations we get 
\begin{equation}
p(x,t) \simeq \frac{1}{2}\sqrt{\frac{r}{\alpha
K_{\alpha}}}t^{\frac{1-\alpha}{2}}\exp\left(-\sqrt{\frac{r}{\alpha
K_{\alpha}}}\left|x\right|t^{\frac{1-\alpha}{2}}\right)\,.
\label{pdfexp1}
\end{equation}
This distribution is evidently non-Gaussian, time-dependent, and has a cusp at $x=0$. For $\alpha=1$,
corresponding to ordinary
Brownian motion, Eq. (\ref{pdfexp1}) tends to stationary steady state, obtained in \cite{EvansMajumdar}.
In Fig. \ref{GSBMexp} we plot the PDF for initially subdiffusive SBM ($\alpha=0.5$) under Poissonian resetting
at different times.
At short times $t<1/r$ the width of PDF is growing, then it starts to decrease, collapsing finally to a vary
narrow function.
The initially subdiffusive motion leads in the course of time to trapping at the origin, 
as was already seen from the behavior of its MSD, which at long times tends to zero as $t^\gamma$ with 
$\gamma = \alpha-1=-0.5$ (orange line in Fig. \ref{GR2exp}). 
In Fig. \ref{GSBMexpsuper} PDF for initially superdiffusive SBM ($\alpha=3$) with Poissonian resetting is
presented at different times $t= 2, 10, 100$ and 1000.
Here the distribution broadens fast and approaches at longer times the scaling form, as given by
Eq.(\ref{pdfexp1}).

\section{Scaled Brownian motion with power-law resetting}

Let us consider the case when the time between successive resets is distributed according to the power law
\begin{equation}
\psi (t) = \frac{\beta/\tau_0}{\left(1+t/\tau_0\right)^{1+\beta}}, \qquad \beta>0
\label{pdfpow}
\end{equation}
with $\tau_0$ assumed to be constant and is set to unity in simulations. The survival probability, according
to Eq. (\ref{surv}), reads
\begin{equation}
\Psi (t) = \left(1+t/\tau_0\right)^{-\beta}\,.
\label{psireset}
\end{equation}
For $\beta>2$ both the first and the second moments of the distribution function of waiting times do exist:
\begin{eqnarray}
\int_0^{\infty}t\psi(t)dt &=& \frac{\tau_0}{\beta-1}, \\
\int_0^{\infty}t^2\psi(t)dt &=&\frac{2\tau_0^2}{\left(\beta-1\right)\left(\beta-2\right)}.
\end{eqnarray}
For $1<\beta<2$ the second moment does not exist while the first moment does. For $\beta<1$ both the first and
the second moments diverge.
As we will see below the parameter $\beta$ has a crucial impact on the behavior of the system. 

In the Laplace domain
\begin{equation}
\tilde{\psi}(s)= \frac{\beta}{\tau_0}\int_0^{\infty}dte^{-ts}\left(1+\frac{t}{\tau_0}\right)^{-1-\beta}.
\end{equation}
Performing the change of the variables $y=s\left(t+\tau_0\right)$ and integrating by parts we get
\begin{equation}\label{psiss}
\tilde{\psi}(s)=1-e^{s\tau_0}\left(s\tau_0\right)^{\beta}\int_{s\tau_0}^{\infty}dye^{-y}y^{-\beta}. 
\end{equation}
For $s\to 0$ and $0<\beta<1$ the integration yields
\begin{equation}
\label{psib01}\tilde{\psi}(s)=1-\Gamma\left(1-\beta\right)\left(s\tau_0\right)^{\beta}+\ldots\, .
\end{equation}
%For $\beta>1$ the last term in Eq.~(\ref{psiss}) may be computed using the additional integration by parts:%\begin{equation}
%\int_{s\tau_0}^{\infty}dye^{-y}y^{-\beta}=\frac{e^{-s\tau_0}\left(s\tau_0\right)^{1-\beta}}{\beta-1}+\frac{1}{1-\beta}\int_{s\tau_0}^{\infty}dye^{-y}y^{1-\beta}
%\end{equation}
%which yields
%\begin{equation}\label{psis1}
%\tilde{\psi}(s)=1-\frac{s\tau_0}{\beta-1}+\frac{e^{s\tau_0}\left(s\tau_0\right)^{\beta}}{\beta-1}\int_{s\tau_0}^{\infty}dye^{-y}y^{1-\beta}+\ldots
%\end{equation}
For $1<\beta<2$ the asymptotic result for $s \to 0$ reads
\begin{equation}
e^{s\tau_0}\int_{s\tau_0}^{\infty}dye^{-y}y^{1-\beta} \to \Gamma\left(2-\beta\right),
\end{equation}
and we get
\begin{equation}
\label{psib12}\tilde{\psi}(s)=1-\frac{s\tau_0}{\beta-1}+\frac{\left(s\tau_0\right)^{\beta}\Gamma\left(2-\beta\right)}{\beta-1}
+ \ldots \, ,
\end{equation}
while for $\beta>2$ %the integration by parts in Eq.~(\ref{psis1}) can be performed to yield
%\begin{equation}
%\int_{s\tau_0}^{\infty}dye^{-y}y^{1-\beta}=\frac{e^{-s\tau_0}\left(s\tau_0\right)^{2-\beta}}{\beta-2}+\frac{1}{\beta-2}\int_{s\tau_0}^{\infty}dye^{-y}y^{2-\beta}
%\end{equation}
%It is straightforward to show that
we get
\begin{equation}
\label{psib2}
\tilde{\psi}(s)=1-\frac{s\tau_0}{\beta-1}+\frac{\left(s\tau_0\right)^{2}}{\left(\beta-1\right)\left(\beta-2\right)}+\ldots
\, .
\end{equation}

%In the following we calculate both analytically and numerically MSD and PDF for three different cases,
$0<\beta<1$, $1<\beta<2$ and $\beta>2$.

\subsection*{$0<\beta< 1$}

\begin{figure}[htbp]
  \centerline{
\includegraphics[width=0.7\columnwidth]{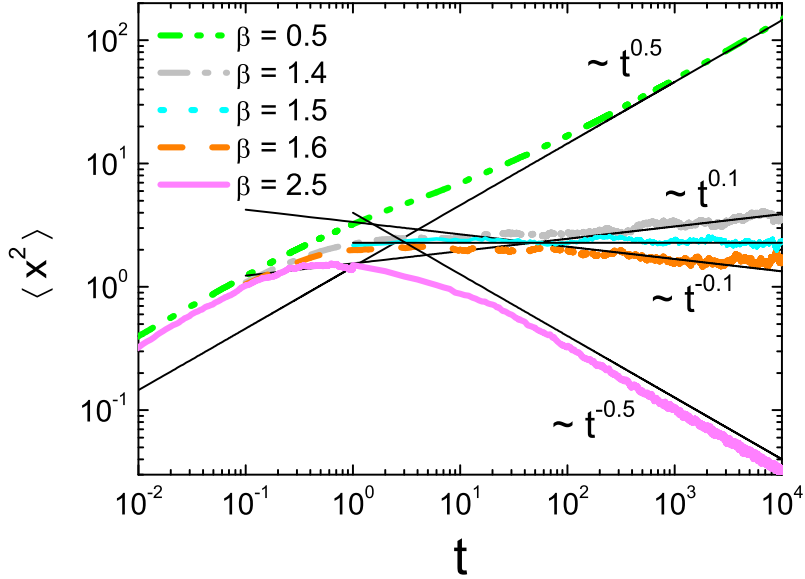}}
\caption{MSD for SBM with the power law resetting, $\alpha=0.5$. This Figure will be referred to several times
in the text.
Thick colored lines correspond to numerical simulations. Thin solid black lines are analytical results. 
For $\beta=0.5$ Eq. (\ref{x2b01}) is used, for $\beta=1.4, 1.5, 1.6$ Eq. (\ref{x2b12}) is used, for
$\beta=2.5$ Eq. (\ref{x2b2}) is used.}
\label{GpowerSBM}
\end{figure}

\paragraph*{Mean squared displacement.} In order to calculate the MSD we use Eq.~(\ref{x21}). The rate of the
resetting events $\kappa(t)$ is given by Eq.~(\ref{psib01}) and Eq.~(\ref{phis}),
\begin{equation}
\tilde{\kappa}(s) \simeq \frac{1}{\Gamma(1-\beta)\tau_0^\beta s^{\beta}} \, ,
\end{equation}
so that 
\begin{equation}
\kappa(t) \simeq \frac{\tau_0^{-\beta}}{\Gamma(\beta)\Gamma(1-\beta)} t^{\beta-1} \,.
\label{phib01}
\end{equation}
At difference with the case of exponential resetting, the rate of resetting events decays with time. 
The MSD for power-law resetting with $0<\beta<1$ can be obtained by inserting Eq. (\ref{phib01}) into Eq.
(\ref{x21}),
\begin{equation}
\left\langle x^2(t)\right\rangle \simeq 2K_{\alpha}t^{\alpha}\left(1-\frac{1}{\alpha
B\left(\alpha,\beta\right)}\right)\, ,
\label{x2b01}
\end{equation}
with $B\left(\alpha,\beta \right)=\int_0^1 dt t^{\alpha-1}\left(1-t\right)^{\beta-1} =
\frac{\Gamma(\alpha)\Gamma(\beta)}{\Gamma(\alpha+\beta)}$ being the Beta function.
We note that $\left(1-\frac{1}{\alpha B \left(\alpha,\beta\right)}\right)< 1$ so that the MSD described by Eq.
(\ref{x2b01}) differs from the MSD for particle performing free diffusive SBM, Eq.~(\ref{x2-AD}),
only by the prefactor: $\left\langle x^2(t)\right\rangle \simeq 2K_{\alpha}^* t^{\alpha}$ with $K_{\alpha}^* <
K_{\alpha}$. The
comparison between analytical and numerical results for power-law resetting is given in Fig. \ref{GpowerSBM}.
Thick, colored lines correspond to
the numerical results, and the dashed lines -- to the theory, showing the nice agreement in the asymptotic
domain.
Initially subdiffusive motion with $\alpha=0.5$ remains
subdiffusive: $\left\langle x^2\right\rangle \sim t^{\alpha}$ (green line corresponding to $\beta=0.5$ in Fig.
\ref{GpowerSBM}). \textcolor{black}{Fig. \ref{GpowerSBM}  gives the overview of all MSD behaviors under power-law resetting discussed in the present paper: 
Other lines in Fig.~\ref{GpowerSBM} show the results 
for the MSD for larger values of $\beta$ as discussed below.}

For $\alpha=1$ the
expression (\ref{x2b01}) yields the normal diffusion regime with different prefactor
\begin{equation}
\left\langle x^2(t)\right\rangle  \simeq 2K_{1}\left(1-\beta\right)t\,.
\label{x2b01obm}
\end{equation}

\paragraph*{Probability density function.}
In order to calculate the PDF we consider the Fourier transform of Eq. (\ref{eqprob}):
\begin{equation}
\hat{p}(k,t) \simeq \int_0^t
dt^{\prime}\kappa(t^{\prime})\Psi(t-t^{\prime})\exp\left(-k^2K_{\alpha}\left(t^{\alpha}-t^{\prime\alpha}\right)\right)\,.
\label{pkt}
\end{equation}

For small $k^2$ (which correspond to large $|x|$ in the far tail of the distribution), $K_{\alpha}k^2
t^{\alpha}\ll 1$, one separates the exponentials containing $t$ and $t'$,
and changes the variable of integration to $\tau=t^{\prime}/t$:
\begin{equation}
\hat{p}(k,t)\simeq\frac{\exp\left(-K_{\alpha}k^2t^{\alpha}\right)}{\Gamma\left(\beta\right)\Gamma\left(1-\beta\right)}\int_0^1
d\tau
\tau^{\beta-1}\left(1-\tau\right)^{-\beta}\exp\left(K_{\alpha}k^2 t^{\alpha}\tau^{\alpha}\right)\,.
\end{equation}
The exponential in the integrand can then be approximated by unity, and the integration yields a 
constant value $B\left(\beta,1-\beta\right)$, so that
\begin{equation}
\hat{p}(k,t)\simeq\exp\left(-K_{\alpha}k^2t^{\alpha}\right)\,.
\end{equation}
The inverse Fourier transform gives the Gaussian behavior of the PDF in its far tail,
\begin{equation}
p(x,t)\simeq\frac{1}{\sqrt{4\pi K_{\alpha}t^{\alpha}}}\exp\left(-\frac{x^2}{4K_{\alpha}t^{\alpha}}\right)\,.
\label{pdfbm1}
\end{equation}
This far tail behavior is universal for power-law resetting.

For $K_{\alpha}k^2 t^{\alpha}\gg 1$ (which corresponds to $x$ in the bulk of the distribution) 
one does not separate the exponentials and uses the approximation
$\tau^{\alpha}\simeq 1-\alpha\left(1-\tau\right)$. Introducing a new variable $\xi=1-\tau$ we find
\begin{equation}
\hat{p}(k,t)\simeq\frac{1}{\Gamma\left(\beta\right)\Gamma\left(1-\beta\right)}\int_0^{1}d\xi
\xi^{-\beta}\left(1-\xi\right)^{\beta-1}e^{-\alpha
K_{\alpha}k^2t^{\alpha}\xi}\, .
\label{eq:in}
\end{equation}
The upper limit of integration can then be shifted to infinity (since the argument of the exponential is very
large and negative),
and then the inverse Fourier transform of this expression may be performed. The result for $x^2 \ll K_{\alpha}
t^{\alpha} $
thus reads:
\begin{equation}
p(x,t)\simeq\frac{1}{\Gamma\left(1-\beta\right)}\frac{1}{\sqrt{4\pi\alpha
K_{\alpha}t^{\alpha}}}\exp\left(-\frac{x^2}{4\alpha
K_{\alpha}t^{\alpha}}\right)U\left(\beta,\beta+\frac{1}{2},\frac{x^2}{4\alpha K_{\alpha}t^{\alpha}}\right)\,.
\label{pdfU}
\end{equation}
Here $U\left(a,b,z\right)$ is the Tricomi confluent hypergeometric function \cite{as}
\begin{equation}
U\left(a,b,z\right)=\frac{1}{\Gamma\left(a\right)}\int_0^{\infty}dt e^{-zt}t^{a-1}\left(1+t\right)^{b-a-1}\,.
\label{Tricomi}
\end{equation}
Using the expansion of $U\left(a,b,z\right)$ for $z\ll 1$ \cite{as}, we get the following asymptotics for
Eq.~(\ref{pdfU}):
\begin{eqnarray}\label{ppp1}
p(x,t)&\simeq&\frac{1}{\sqrt{4\alpha
K_{\alpha}t^{\alpha}}\pi^{2}}\sin\left(\pi\beta\right)\Gamma\left(1/2-\beta\right)\Gamma\left(\beta\right)\,\,\,,
0\le\beta < 1/2, \\
p(x,t)&\simeq& \frac{\Gamma\left(\beta-1/2\right)}{\left(4\alpha
K_{\alpha}t^{\alpha}\right)^{1-\beta}}\frac{\sin\left(\pi\beta\right)}{\pi^{3/2}\left|x\right|^{2\beta-1}}\,\,\,,
1/2 < \beta\le 1. \label{ppp2}
\end{eqnarray}
This change of the behavior can be anticipated from the form of the integral defining the Tricomi function,
Eq.(\ref{Tricomi}),
since for $\beta > 1/2$ the integral diverges at the upper limit for $z=0$, while for $\beta < 1/2$ it
converges at the upper limit also without the regularizing exponential
depending on $x$, so that the distribution at small $x$ develops a flat top. The transition involving
logarithmic corrections is not captured by the asymptotic expansions.

For $\alpha=1$ the PDF behaves as that for the ordinary Brownian motion with power-law resetting with
$0<\beta<1$ \cite{NagarGupta}.

\begin{figure}[htbp]
  \centerline{
\includegraphics[width=0.7\columnwidth]{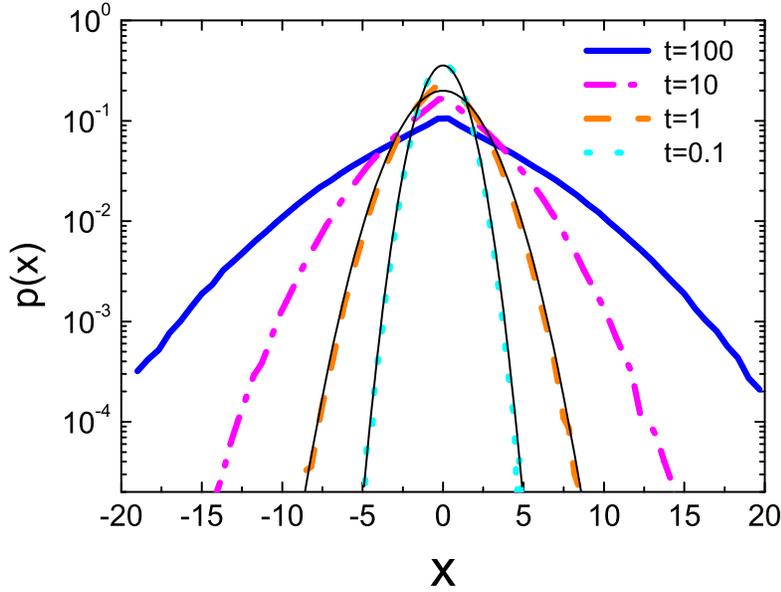}
}
\caption{PDF for SBM with power-law resetting, $\beta=0.25$, $\alpha=0.5$ at times $t=0.1, 1, 10$ and 100.
Thin black solid lines show the Gaussians
$\exp\left(-x^2/4K_{\alpha}t^{\alpha}\right)/\sqrt{ 4 \pi K_{\alpha}t^{\alpha}}$, 
Eq.(\ref{pdfbm1}). \label{GSBM025}}
\end{figure}

In Fig.~\ref{GSBM025} we show the results of numerical simulations for $\beta=0.25$ at shorter times, which is
indeed
well fitted with Gaussian function, Eq. (\ref{pdfbm1}). At variance with the case of subdiffusive SBM with
exponential resetting, the width of the probability distribution grows.

\begin{figure}[htbp]
  \centerline{
\includegraphics[width=0.7\columnwidth]{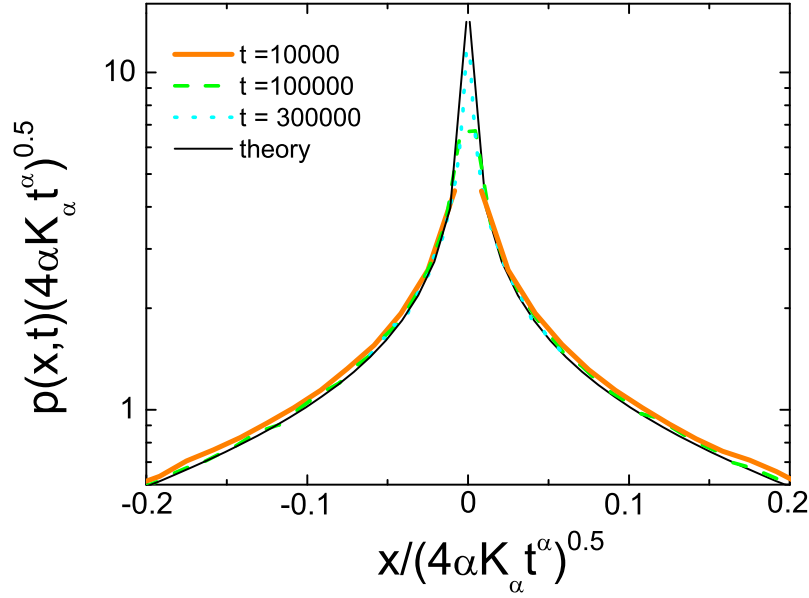}
}
\caption{PDF for SBM with power-law resetting, $\beta=0.75$, $\alpha=0.5$ at long times. The thin black solid
line is Eq. (\ref{ppp2}). }
\label{GSBM075}
\end{figure}

In Fig.~\ref{GSBM075} the numerical results for PDF for SBM with power-law resetting, $\beta=0.75 > 1/2$, is
given. In rescaled variables
$p(x,t)\sqrt{4\alpha K_{\alpha}t^{\alpha}}$ versus $x/\sqrt{4\alpha K_{\alpha}t^{\alpha}}$ the distribution
functions
at different times collapse, and show a nice agreement with the analytical solution, Eq. (\ref{ppp2}).

\subsection*{$1<\beta<2$ }%\label{Sec:inter}}

\paragraph*{Mean squared displacement.} The calculation leading to the MSD is similar to the case $0<\beta<1$.
We use again Eq. (\ref{x21}) but with a constant rate of resetting
\begin{equation}\label{phib12t}
\kappa(t) \simeq \kappa =\frac{\beta-1}{\tau_0} .
\end{equation}
%which in Laplace domain corresponds to
%\begin{equation}
%\tilde{\kappa}(s)=\frac{\beta-1}{s\tau_0}\,.
%\label{phib12}
%\end{equation}
Plugging Eq. (\ref{phib12t}) into Eq. (\ref{x21}) results in
\begin{equation} %\label{C1}
\left\langle x^2(t)\right\rangle \simeq \frac{2K_{\alpha}\left(\beta-1\right)}{\tau_0}\left[\int_0^t
dt^{\prime}\Psi\left(t-t^{\prime}\right)t^{\alpha}-\int_0^t
dt^{\prime}\Psi\left(t-t^{\prime}\right)t^{\prime\alpha}\right]
\label{MsdLargeB}
\end{equation}
with $\Psi(t)$ given by Eq. (\ref{psireset}). The integration in the first term is straightforward.
The second term has a form of a convolution, and the integral can be evaluated in the Laplace domain using
\begin{equation}
\tilde{\Psi}\left(s\right)\simeq\frac{\tau_0}{\beta-1}-\frac{\Gamma (2-\beta)s^{\beta-1}\tau_0^{\beta}}{\beta
- 1} ,
\label{Psis12}
\end{equation}
as follows from Eq.~(\ref{psib12}) and Eq. (\ref{survivalLaplace}), so that
\begin{equation}\label{C3}
\mathcal{L}\left\{\int_0^t
dt^{\prime}\Psi\left(t-t^{\prime}\right)t^{\prime\alpha}\right\}\simeq\frac{\Gamma\left(\alpha+1\right)}{s^{\alpha+1}}\left(\frac{\tau_0}{\beta-1}-\frac{\Gamma
(2-\beta)s^{\beta-1}\tau_0^{\beta}}{\beta - 1}\right) .
\end{equation}
By taking the inverse Laplace transform of Eq.~(\ref{C3}), we obtain
\begin{equation}\label{C4}
\int_0^t
dt^{\prime}\Psi\left(t-t^{\prime}\right)t^{\prime\alpha}\simeq\frac{\tau_0}{\beta-1}\left(t^{\alpha}-\frac{\Gamma\left(2-\beta\right)\Gamma\left(1+\alpha\right)t^{1-\beta+\alpha}\tau_0^{\beta-1}}{\Gamma\left(\alpha-\beta+2\right)}\right).
\end{equation}
The final result reads:
\begin{equation}
\left\langle x^2(t)\right\rangle \simeq 2K_{\alpha}t^{1+\alpha-\beta}\tau_0^{\beta-1}\left(\alpha
B\left(\alpha,2-\beta\right)-1\right).
\label{x2b12}
\end{equation}
The system with $1<\beta<2$ demonstrates a very rich behavior. The exponent of time-dependence of MSD
decreases by the amount $\beta-1$
compared to the free motion. This amount changes from 0 for $\beta=1$ to 1 for $\beta=2$. For $\beta<1+\alpha$
the
MSD grows with time, and in the opposite case $\beta>1+\alpha$ it decays at long times, in which case
the particles are unable to move far away from the origin. In the case of superdiffusion the motion of
particles either remains
superdiffusive, can tend to ordinary diffusion, or become subdiffusive. In the case of subdiffusion the motion
can either slow down or get suppressed.
For $\beta=1+\alpha$ the MSD stagnates. This is however a very intriguing situation, since, as we proceed to
show, the
stagnation of the MSD does not imply the existence of the NESS.

\begin{figure}[htbp]
  \centerline{
\includegraphics[width=0.6\columnwidth]{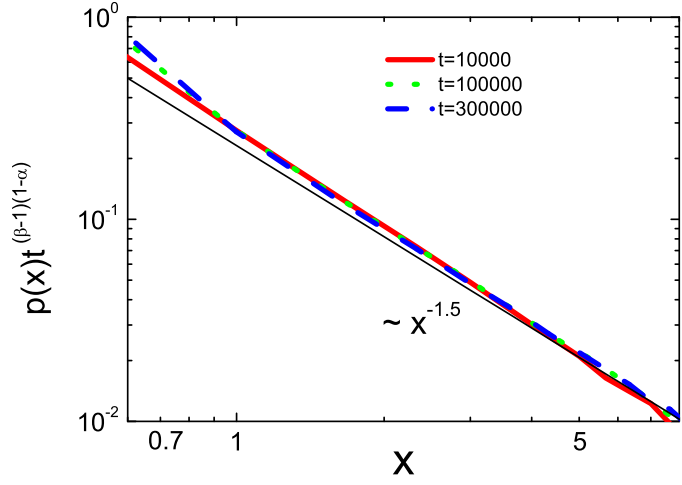}
}
\caption{The PDF for SBM with power-law resetting, $\beta=1.25$, $\alpha=0.5$ rescaled according to
Eq.~(\ref{pdfb1}). Shown is $p(x)t^{(\beta-1)(1-\alpha)}$ as a function of $x$ for $x >0$.
Thin black line has the slope $-3/2$ as following from Eq.(\ref{pdfb1}). \label{GSBM125}}
\end{figure} 

The comparison between numerical and analytical results for this case is also presented in
Fig.~\ref{GpowerSBM} above. For $1+\alpha>\beta$ the system remains subdiffusive but
with lower exponent (gray line corresponding to $\beta=1.4$ in Fig. \ref{GpowerSBM}), for $\beta=1+\alpha$ the
MSD stagnates as depicted in Fig.
\ref{GpowerSBM} (blue line corresponding to $\beta=1.5$ in Fig. \ref{GpowerSBM}). For $1+\alpha<\beta$ the MSD
tends to zero (orange line
corresponding to $\beta=1.6$ in Fig. \ref{GpowerSBM}). %It means that the particles are trapped in the origin.% The motion between two resetting event is too slow and the resetting events occur too frequently, so the
particle can not move away from the origin
%for a significant distance.
%In such a way, for subdiffusive SBM small changes in the power-law index of the waiting time distribution of
resetting events can drastically affect
%the behavior of the system, changing it from subdiffusion in the case of $1<\beta<1+\alpha$ to stagnant for
$\beta=1+\alpha$, and leading to trapping
%of the particle at the origin for $\beta>1+\alpha$.

In the case of the ordinary Brownian motion with resetting the particle performs subdiffusive motion
\begin{equation}
\left\langle x^2(t)\right\rangle  \simeq 2K_{1}\tau_0^{\beta-1}t^{2-\beta}\frac{\beta-1}{2-\beta} .
\label{x2b12obm}
\end{equation}
This expression can be directly obtained from Eq. (\ref{x2b12}) by taking $\alpha=1$.

\paragraph*{Probability density function.} 

Inserting the expressions for $\kappa$, Eq.(\ref{phib12t}), and $\Psi(t)$, Eq. (\ref{psireset}), into Eq.
(\ref{eqprob}) we get in the time domain
\begin{equation}
 p\left(x,t\right)\simeq\frac{\beta-1}{\tau_0}\frac{1}{\sqrt{4\pi K_{\alpha}}}\int_0^t
dt^{\prime}\left(1+\frac{t-t^{\prime}}{\tau_0}\right)^{-\beta}\frac{1}{\sqrt{t^{\alpha}-t^{\prime\alpha}}}\exp\left(-\frac{x^2}{4K_{\alpha}\left(t^{\alpha}-t^{\prime\alpha}\right)}\right)\,.
\end{equation}
Now we assume $t\gg\tau_0$ and change the variable of integration to $y \simeq
\left(1-\left(t^{\prime}/t\right)^{\alpha}\right)^{-1}$:
\begin{equation}
p(x,t) \simeq \frac{t^{1-\beta-\frac{\alpha}{2}}\tau_0^{\beta}}{\alpha}\int_1^{\infty} dy
y^{-\frac{1}{2}-\frac{1}{\alpha}}\left(y-1\right)^{\frac{1}{\alpha}-1}\left(1-\left(1-\frac{1}{y}\right)^{\frac{1}{\alpha}}\right)^{-\beta}\exp\left(-\frac{x^2y}{4K_{\alpha}t^{\alpha}}\right)\,.
\end{equation}
For intermediate values of $x$ (i.e. in the core of the PDF, but not very close to its mode) the integral is
dominated by large values of $y$,
where we can make the approximations $\left(y-1\right)^{\frac{1}{\alpha}-1}\approx y^{\frac{1}{\alpha}-1}$ and$\left(1-\left(1-\frac{1}{y}\right)^{\frac{1}{\alpha}}\right)^{-\beta}\approx \alpha^{\beta}y^{\beta}$. The
expression is now simplified to
\begin{equation}
p(x,t)\simeq t^{1-\beta-\frac{\alpha}{2}}\alpha^{\beta-1}\tau_0^{\beta}\int_1^{\infty} dy
y^{\beta-\frac32}\exp\left(-\frac{x^2y}{4K_{\alpha}t^{\alpha}}\right)\,.
\end{equation}
The lower bound of integration may be safely shifted to zero, so that  
\begin{equation}
p(x,t) \simeq
\alpha^{\beta-1}\tau_0^{\beta}\left(4K_{\alpha}\right)^{\beta-\frac12}\Gamma\left(\beta-\frac12\right)t^{\left(\beta-1\right)\left(\alpha-1\right)}x^{1-2\beta}
.
\end{equation}
Omitting the prefactors we get 
\begin{equation}
p\left(x,t\right)\sim x^{1-2\beta}t^{(1-\beta)(1-\alpha)}\,.
\label{pdfb1}
\end{equation}
The distribution can be put into a scaling form $p(x,t) = t^{-\gamma} f(x/t^\gamma)$ with $f(z)=z^{1-2\beta}$,
so that $\gamma =(\alpha -1)/2$.
This scaling form is shown in Fig.~\ref{GSBM125}.  For
ordinary Brownian motion with power-law resetting and $\beta>1$ the steady state $p\left(x,t\right)\sim
x^{1-2\beta}$ is recovered \cite{NagarGupta}.

Now we return to a balanced situation $\beta=1+\alpha$ when the MSD stagnates, and see that the bulk of the
distribution stays time-dependent: the
stagnation of the MSD is due to the compensation effect between the narrowing central peak and growing tail,
which is a quite peculiar situation to no
extent representing a NESS.

At long times for parameter values $\beta=1.25$, $\alpha=0.5$ the PDF indeed follows the scaling predicted by
Eq.(\ref{pdfb1}) and has an asymptotics
$p(x)\simeq x^{-3/2}$ as shown in Fig.~\ref{GSBM125}.

\subsection*{$\beta>2$}

\paragraph*{Mean squared displacement.} The MSD can be obtained similar to the previous case, $1<\beta<2$. 
Inserting Eq.~(\ref{psib2}) into Eq. (\ref{survivalLaplace}), we get for the Laplace transform of the survival
probability
\begin{equation}
\tilde{\Psi}\left(s\right)\simeq\frac{\tau_0}{\beta-1}-\frac{s\tau_0^2}{\left(\beta-1\right)\left(\beta-2\right)}\,.
\label{Psis2}
\end{equation}
Now we can calculate the MSD using Eq.(\ref{MsdLargeB}). 
The final form of the MSD reads:
\begin{equation}
\left\langle x^2(t)\right\rangle  \simeq \frac{2\alpha K_{\alpha}\tau_0}{\beta-2} t^{\alpha-1} .
\label{x2b2}
\end{equation}
This behavior resembles MSD for Poissonian resetting (Eq. \ref{x2exp}). Note that the exponent in
time-dependence of MSD, Eq. (\ref{x2b2}),
ceases to depend on the exponent $\beta$ of the waiting time distribution, the dependence on $\beta$ stays
only in the prefactor. A nice agreement
with the numerical simulation, shown as magenta line corresponding to $\beta=2.5$ in Fig. \ref{GpowerSBM} is
observed.

In the case of ordinary Brownian motion ($\alpha=1$) the MSD stagnates
\begin{equation}
\left\langle x^2(t)\right\rangle  \simeq \frac{2K_{1}\tau_0}{\beta-2}\,.
\label{x2b2obm}
\end{equation}

\paragraph*{Probability density function.} The asymptotics for PDF has the same form, Eq. (\ref{pdfb1}), as
for $1<\beta<2$.

\section{Conclusions}

In the present work we discussed the MSD and the PDF for particles performing scaled Brownian motion with time-dependent diffusion 
coefficient $D(t)\sim t^{\alpha-1}$ under resetting in a non-renewal case, when the position of a particle is returned to the origin upon resetting, while the diffusion coefficient 
(changing with time) remains unaffected by the resetting events. The distribution of waiting times between two successive resetting events 
was either exponential $\psi (t) \sim {e^{ - rt}}$ or followed a power-law $\psi (t) \sim t^{-1-\beta}$. To the best of our knowledge, this is the
first exhaustive study of a stochastic process which is not rejuvenated at a resetting event.

For $\beta<1$ the power-law exponent of MSD is not affected by resetting,  $\left\langle x^2 \right\rangle \simeq t^{\alpha}$, but
only changes the prefactor. For $1<\beta<2$ the MSD scales as $\left\langle x^2 \right\rangle \simeq t^{1+\alpha-\beta}$, and the behavior of the
system is determined by the interplay of the exponents $\alpha$ and $\beta$, so that the particle's motion is either slowed down compared to the free SBM or completely suppressed. 
Interestingly enough, the compensated case when the MSD stagnates does not correspond to a stationary state, since the PDF still changes with time. 

The cases of Poissonian resetting and of power-law resetting with $\beta>2$ show strong similarities in the behavior of the MSD: in both cases it scales as 
$\left\langle x^2 \right\rangle \simeq t^{\alpha-1}$. This means that such resetting always decreases the exponent of the MSD by unity, so that for
$\alpha > 2$ the initially superdiffusive motion remains superdiffusive, for $1<\alpha<2$ superdiffusion tends to subdiffusion, and subdiffusive
motion with $\alpha<1$ becomes completely suppressed: the particles get trapped in the vicinity of the starting point. 

Since the SBM for $0< \alpha < 1$ shows the same aging properties of MSD as the CTRW, the very same behavior could be anticipated for resetting of CTRW provided the 
resetting events do not rejuvenate the waiting times.

The PDF of the particle's position for non-renewal resetting with exponential waiting time PDF is non-stationary but always shows simple 
two-sided exponential (Laplace) shape. In the case of power-law resetting waiting time PDF with very slow decay ($\beta < 1/2$) the PDF of positions does not show any universal scaling 
in the body and possesses Gaussian tails. In all other cases it tends to universal forms which differ in their time-dependent prefactor for $1/2 < \beta < 1$, and for $\beta > 1$. 
The behavior of the MSD and of the PDF in the bulk is presented in Table \ref{TableI}. 

\begin{table}[h]
\caption{Asymptotic behavior of the MSD and of the PDF in the intermediate domain of $x$ for power-law resetting. \label{TableI}}
\centering
\begin{tabular}{ | l | l | l |  l | l | }
\hline
 & $0<\beta<1/2$ & $1/2<\beta<1$ & $1<\beta<2$ & $\beta>2$ \\ \hline
MSD & $\sim t^{\alpha}$ & $\sim t^{\alpha}$ & $\sim t^{\alpha+1-\beta}$ & $\sim t^{\alpha-1}$ \\
PDF & $\begin{array}{cc}
       \mbox{flat top, Gaussian tail}
      \end{array}$
 & $\sim t^{\alpha(\beta-1)}\left|x\right|^{1-2\beta}$ & $\sim t^{(1-\beta)(1-\alpha)}\left|x\right|^{1-2\beta}$ & $\sim
t^{(1-\beta)(1-\alpha)}\left|x\right|^{1-2\beta}$ \\
\hline
\end{tabular}
\end{table}

\color{black}

These results should be confronted with the ones for the situation when the transport process is rejuvenated under resetting, and the whole process
is a renewal one, as discussed in detail in the other work of these series, Ref. \cite{Anna0}. The behavior observed in such a renewal process significantly differs from the results discussed above. 
Here the behavior of the MSD is as follows: For exponential resetting and power-law resetting with $\beta>1+\alpha$ the  MSD at long times stagnates.
For $\beta<1$ the time dependence of the MSD remains the same as in the case of free scaled Brownian motion, albeit with different prefactors. 
In the intermediate domain $1<\beta<1+\alpha$ we obtain $\left\langle x^2\right\rangle\sim t^{1+\alpha-\beta}$, so that the behavior of the MSD is defined by the interplay of 
the parameters $\alpha$ and $\beta$. 

Turning to the behavior of the PDF we state that in the case of the exponential resetting the PDF tends to a steady state with stretched or squeezed exponential tail $p(x,t)\simeq \exp\left(-\gamma
|x|^{\frac{2}{\alpha+1}}\right)$. For the power-law resetting with $\beta>1$ the PDF also attains a time-independent form, now $p(x,t)\sim x^{-1-\frac{2\beta}{\alpha}+\frac{2}{\alpha}}$. 
We note that for $\beta>1+\alpha$ both the MSD and the PDF tend to the stationary state, while for $1<\beta<1+\alpha$ only the PDF in the bulk is stationary but the
MSD grows continuously with time. For $\beta<1$ the behavior of the PDF depends on the relation between the
exponents $\beta$ and $\alpha$. For $\beta > 1 - \alpha/2$ the $x$-dependence of the PDF for $\sqrt{4 K_\alpha \tau_0^\alpha} \ll |x| \ll \sqrt{4 K_\alpha t^\alpha}$ is the same as in the previous case,
but now the time-dependence also appears: $p(x,t)\sim t^{\beta-1}\left|x\right|^{-1-\frac{2\beta}{\alpha}+\frac{2}{\alpha}}$. For long times this intermediate domain covers
practically the whole bulk of the distribution. For $\beta < 1 - \alpha/2$ the PDF in the center of the distribution is flat, with a Gaussian tail at $x \gg \sqrt{4 K_\alpha t^\alpha}$. 
The results for the MSD and the PDF are presented in the Table \ref{TableII}.

\begin{table}[h!]
\caption{MSD and PDF for the renewal power-law resetting \label{TableII}}
\centering
\begin{tabular}{ | l | l | l | l |  l | }
\hline
 & $0<\beta<1-\alpha/2$ & $1-\alpha/2 < \beta < 1$ & $1<\beta<1+\alpha$ & $\beta>1+\alpha$ \\ \hline
MSD & $\sim t^{\alpha}$ & $\sim t^{\alpha}$ & $\sim t^{\alpha+1-\beta}$ & stagnates \\
PDF &flat top, Gaussian tail & $\sim t^{\beta-1}\left|x\right|^{-1-2\beta/\alpha+2/\alpha}$ & $\sim \left|x\right|^{-1-2\beta/\alpha+2/\alpha}$ & $\sim
\left|x\right|^{-1-2\beta/\alpha+2/\alpha}$ \\
\hline
\end{tabular}
\end{table}

The comparison of the results for renewal and non-renewal variants of the same process shows that erasing or retaining the memory in transport process 
is crucial for the features of the overall dynamics, which is the main physical consequence drawn in the present work. To the best of our knowledge, the SBM is the only process for which such a comparison 
was performed. 

\color{black}

\section{Acknowledgements} AVC is indebted to D. Boyer for fruitful discussions which initiated this work, and
acknowledges the financial support by the Deutsche Forschungsgemeinschaft within the project ME1535/6-1.\\


\begin{thebibliography}{99}

\bibitem{Gerstner} W. Gerstner and W. M. Kistler, \textit{Spiking neuron models: Single neurons, populations,
plasticity}, Cambridge Univ. Press, 2002
\bibitem{Manrubia99} S. C. Manrubia and D. H. Zanette, Phys. Rev. E \textbf{59}, 4945 (1999).
\bibitem{EvansMajumdar} M. R. Evans and S. N. Majumdar, Phys. Rev. Lett. \textbf{106}, 160601 (2011).
\bibitem{levy1} L. Ku\'smierz, S. N. Majumdar, S. Sabhapandit, and G. Schehr, Phys. Rev. Lett. \textbf{113},
220602 (2014).
\bibitem{levy2} L. Ku\'smierz, E. Gudowska-Nowak. Phys. Rev. E \textbf{92}, 052127 (2015).
\bibitem{MSS2015} S. N. Majumdar, S. Sabhapandit, and G. Schehr, Phys. Rev. E \textbf{91}, 052131 (2015).
\bibitem{high12} M.R. Evans and S.N. Majumdar. J. Phys. A: Math. Theor. \textbf{47}, 285001 (2014).  
\bibitem{CS2015}  C. Christou and A. Schadschneider, J. Phys. A: Math. Theor. \textbf{48}, 285003 (2015).
\bibitem{ArnabPal} A. Pal and  V. V. Prasad. Phys. Rev. E \textbf{99}, 032123 (2019).
\bibitem{P2015}  A. Pal, Phys. Rev. E \textbf{91}, 012113 (2015).
\textcolor{black}{\bibitem{expot} S. Ray, D. Mondal, Sh. Reuveni. https://arxiv.org/abs/1811.08239.}
\bibitem{EM2011}  M. R. Evans and S. N. Majumdar, J. Phys. A: Math. Theor. \textbf{44}, 435001 (2011).
\bibitem{BS2014}  D. Boyer and C. Solis-Salas, Phys. Rev. Lett. \textbf{112}, 240601 (2014).
\bibitem{MSS2015a}  S. N. Majumdar, S. Sabhapandit, and G. Schehr, Phys. Rev. E \textbf{92}, 052126 (2015).
\textcolor{black}{\bibitem{nonstatic} M.A. dos Santos, Physics, \textbf{1}, 40 (2019).
\bibitem{brm2} R Falcao and M R Evans, J. Stat. Mech. 023204 (2017).
\bibitem{reacdiff} G.J. Lapeyre, M. Dentz, Phys. Chem. Chem. Phys., \textbf{19}, 18863 (2017).
\bibitem{branch1} I. Eliazar Europhys. Lett., 120, 60008 (2018).
\bibitem{branch2} A. Pal, I. Eliazar, S. Reuveni, Phys. Rev. Lett., 122, 020602 (2019).}
\bibitem{Rosemary} R.J. Harris and H. Touchette, J. Phys. A: Math. Theor. \textbf{50} 10LT01 (2017)
\bibitem{res2016} S. Eule and J. J. Metzger, New J. Phys. \textbf{18}, 033006 (2016).
\bibitem{palrt} A. Pal, A. Kundu and M. R. Evans, J. Phys. A: Math. Theor. \textbf{49}, 225001 (2016).
%\bibitem{shlomi2017} A. Pal and S. Reuveni. Phys. Rev. Lett. \textbf{118}, 030603 (2017).
%\bibitem{shlomi2016} S. Reuveni. Phys. Rev. Lett. \textbf{116}, 170601 (2016).
\bibitem{NagarGupta} A. Nagar and S. Gupta, Phys. Rev. E \textbf{93}, 060102 (2016).
\bibitem{coagdiff} X. Durang, M. Henkel, and H. Park, J. Phys. A \textbf{47}, 045002 (2014).
\bibitem{boyer2017} D. Boyer, M.R. Evans, S.N. Majumdar. J. Stat. Mech. 023208 (2017).
\bibitem{MV2013}  M. Montero and J. Villarroel, Phys. Rev. E \textbf{87}, 012116 (2013).
\bibitem{MC2016}  V. M$\rm\acute{e}$ndez and D. Campos, Phys. Rev. E \textbf{93}, 022106 (2016).
\bibitem{Sh2017} V.P. Shkilev, Phys. Rev. E \textbf{96}, 012126 (2017).
\bibitem{ctrw} M. Montero, A. Mas-Puigdell\'osas, J. Villarroel. Eur. Phys. J. B \textbf{90}, 176 (2017).
\bibitem{sokolovsub} Y. Meroz, I.~M. Sokolov, Phys. Rep. \textbf{573}, 1 (2015).
\bibitem{sokbook} J. Klafter and I.M. Sokolov, First Steps in Random Walks: From Tools to Applications. Oxford
University Press, New York, USA (2011).
\bibitem{sokolovSM} I.~M. Sokolov, Soft Matter \textbf{8}, 9043 (2012).
\bibitem{MetzlerREVIEW} R. Metzler and J. Klafter, Phys. Rep. \textbf{339}, 1 (2000).
\bibitem{MetzlerPT} E. Barkai, Y. Garini, and R. Metzler, Phys. Today \textbf{65}, 29-35 (2012).
\bibitem{Bouchaudrev} J.-P. Bouchaud and A. Georges, Phys. Rep. \textbf{195}, 127 (1990).
\bibitem{zaburdaev} V. Zaburdaev, S. Denisov, and J. Klafter, Rev. Mod. Phys. \textbf{87}, 483 (2015).
\bibitem{hoefling} F. H\"ofling and T. Franosch, Rep. Prog. Phys. \textbf{76}, 046602 (2013).
\bibitem{ultraslow} A.S. Bodrova, A. V. Chechkin, A. G. Cherstvy, and R. Metzler, New J. Phys. \textbf{17},
063038 (2015).
\bibitem{underdamped1} A.S. Bodrova, A. V. Chechkin, A. G. Cherstvy, H. Safdari, I. M. Sokolov, and R.
Metzler, Sci. Rep. \textbf{6}, 30520 (2016).
\bibitem{underdamped2} H. Safdari, A. G. Cherstvy, A.S. Bodrova, A. V. Chechkin, R. Metzler. Phys. Rev. E
\textbf{95}, 012120 (2017).
\bibitem{batchelor} G. K. Batchelor, Math. Proc. Cambridge Phil. Soc. \textbf{48}, 345 (1952).
\bibitem{Richardson} L. F. Richardson, Proc. R. Soc. London Sect. A \textbf{110}, 709 (1926).
\bibitem{Monin} A. S. Monin and A. M. Yaglom, \textit{Statistical Fluid Mechanics}, Vol. 2, MIT Press,
Cambridge, MA, 1987.
\bibitem{Shlesinger} M.F. Shlesinger, B.J. West, and J. Klafter, Phys. Rev. Lett. \textbf{58}, 110 (1987).
\bibitem{saxton} M. J. Saxton, Biophys. J. \textbf{81}, 2226 (2001).
\bibitem{novikov1} D. S. Novikov, E. Fieremans, J. H. Jensen, J. A. Helpern, Nat. Phys. \textbf{7}, 508
(2011).
\bibitem{novikov2} D.~S. Novikov, J.~H. Jensen, J. A. Helpern, E. Fieremans, Proc. Natl. Acad. Sci. USA
\textbf{111}, 5088 (2014).
\bibitem{molini} A. Molini, P. Talkner, G.~G. Katul, A. Porporato, Physica A \textbf{390}, 1841 (2011).
\bibitem{snow} D. De Walle, A. Rango, Principles of Snow Hydrology, Cambridge University Press, Cambridge, UK
(2008).
\bibitem{brilbook} N.~V. Brilliantov and T. P\"oschel, Kinetic Theory of Granular Gases, Oxford University
Press, Oxford (2004).
\bibitem{ggg1} {\it Granular Gases}, edited by T. Poeschel and S. Luding, Lecture Notes in Physics Vol.
\textbf{564}
(Springer, Berlin, 2001).
\bibitem{ggg2} {\it Granular Gas Dynamics}, edited by T. Poeschel and N. V. Brilliantov, Lecture Notes
in Physics Vol. \textbf{624} (Springer, Berlin, 2003).
\bibitem{megg} A.S. Bodrova, A. V. Chechkin, A. G. Cherstvy, and R. Metzler, Phys. Chem. Chem. Phys.
\textbf{17}, 21791 (2015).
\bibitem{LimSBM} S.~C. Lim and S.~V. Muniandy, Phys. Rev. E \textbf{66}, 021114 (2002).
\bibitem{Hadise} H. Safdari, A. G. Cherstvy, A. V. Chechkin, F. Thiel, I. M. Sokolov, R. Metzler, J. Phys. A:
Math. Theor. \textbf{48}, 375002 (2015).
\bibitem{ThielSok} F. Thiel and I.~M. Sokolov, Phys. Rev. E \textbf{89}, 012115 (2014).
\bibitem{Anna0} A.S. Bodrova, A.V. Chechkin, I.M. Sokolov, submitted.
\bibitem{as} M. Abramowitz and I. A. Stegun. Handbook of Mathematical Functions. Dover, Publications, New
York, NY (1965).
\bibitem{MetzlerSBM} J.-H. Jeon, A.~V. Chechkin, and R. Metzler, Phys. Chem. Chem. Phys. \textbf{16}, 15811
(2014).
\bibitem{Oshanin} S.N. Majumdar and G. Oshanin J. Phys. A: Math. Theor. \textbf{51} 435001 (2018).


\end{thebibliography}
\end{document}